\documentstyle[aps,prl,twocolumn,floats,psfig,epsf]{revtex}

\begin{document}
\draft
\twocolumn[\hsize\textwidth\columnwidth\hsize\csname@twocolumnfalse\endcsname
\title{
The critical behavior of frustrated spin models with noncollinear order
}
\author{Andrea Pelissetto$\,^1$, Paolo Rossi$\,^2$, 
and Ettore Vicari$\,^2$ }
\address{$^1$
Dipartimento di Fisica dell'Universit\`a di Roma I
and I.N.F.N., I-00185 Roma, Italy 
}
\address{$^2$ 
Dipartimento di Fisica dell'Universit\`a 
and I.N.F.N., 
Via Buonarroti 2, I-56127 Pisa, Italy.
\\
{\bf e-mail: \rm
{\tt Andrea.Pelissetto@roma1.infn.it},
{\tt rossi@df.unipi.it}, 
{\tt vicari@df.unipi.it}
}}

\date{\today}
\maketitle

\begin{abstract}
We study the critical behavior of frustrated spin models with noncollinear 
order, including stacked triangular antiferromagnets and helimagnets. 
For this purpose we compute the field-theoretic expansions at fixed
dimension to six loops and determine their large-order behavior. For the 
physically relevant cases of two and three components, we show the 
existence of a new stable fixed point that corresponds to the 
conjectured chiral universality class.  This contradicts previous three-loop
field-theoretical results but is in agreement with experiments.
\end{abstract}


]


The critical behavior of frustrated spin systems with noncollinear or 
canted order
has been the object of intensive
theoretical and experimental studies
(see, e.g., Ref. \cite{Kawamura-98}).  In spite of these efforts, 
the critical behavior of these systems is still unclear,
field-theoretic (FT) renormalization-group (RG) methods, Monte Carlo simulations
and experiments obtaining different results.

In physical magnets noncollinear order is due to frustration that may arise
either because of the special geometry 
of the lattice, or from the competition of different kinds of interactions.
Typical examples of systems of the first type are three-dimensional
stacked triangular antiferromagnets (STA), 
where magnetic ions are located at each site of 
a three-dimensional stacked triangular lattice.
Examples are ${\rm ABX}_3$-type compounds,
where A denotes elements such as Cs and Rb, B stands for magnetic
ions such as Mn, Cu, Ni, and Co, and X for halogens as Cl, Br, and
I. They may  be modeled by using short-ranged Hamiltonians 
for $N$-component spin variables 
defined on a stacked triangular lattice as
\begin{equation}
{\cal H}_{\rm STA} = 
     - J\,\sum_{\langle ij\rangle_{xy}}  \vec{s}_i \cdot \vec{s}_j -
       J'\,\sum_{\langle ij\rangle_z}  \vec{s}_i \cdot \vec{s}_j,
\label{latticeH}
\end{equation}
where $J<0$, the first sum is over nearest-neighbor pairs within
triangular layers ($xy$ planes), and the second one is over 
orthogonal interlayer nearest neighbors.
In these spin systems the Hamiltonian is minimized by 
noncollinear configurations, 
showing a 120$^o$ spin structure.
Frustration is partially released by mutual spin canting,
and the degeneracy of the ground-state is limited to global O($N$) 
spin rotations and reflections. 
As a consequence,  at criticality there is 
a breakdown of the symmetry from O($N$) in the high-temperature phase
to O($N-2$) in the low-temperature phase, 
implying a matrix-like order parameter.
Frustration due to the competition of interactions may be realized in helimagnets where
a magnetic spiral is formed along a certain direction of the lattice
(see, e.g., Ref.~\cite{Kawamura-98}). 
The rare-earth metals Ho, Dy and Tb provide
examples of such systems.

The critical behavior of two- and three-component
frustrated spin models with noncollinear order
is controversial.
Many experiments (see, e.g., Ref.~\cite{Kawamura-98})
are consistent with a second-order phase transition belonging to a new (chiral) 
universality class.  
This is partially supported by Monte Carlo simulations 
(see, e.g., Ref.~\cite{Kawamura-98} and references
therein). On the other hand, 
three-loop perturbative calculations at fixed dimension $d=3$~\cite{AS-94}
and within the framework of the $\epsilon$-expansion~\cite{ASV-95}
indicate a first-order transition, since no stable chiral fixed points are
found for $N=2$ and $N=3$. 
These three-loop analyses show the presence of a stable chiral fixed point
only for $N>N_c$ with $N_c>3$: 
$N_c=3.91$~\cite{AS-94} and $N_c=3.39$~\cite{ASV-95}.

To explain these contradictory results it has been suggested that these
systems undergo
weak first-order transitions, that effectively appear as second-order 
ones in numerical and experimental works. This 
hypothesis has been  supported by studies
based on approximate solutions of 
the Wilson RG equations~\cite{ERG},
and by Monte Carlo investigations~\cite{MC}  
of modified lattice spin systems which, according to 
general universality ideas,
should belong to the same universality class of the 
Hamiltonian (\ref{latticeH}),
and which show a first-order transition.

For larger values of $N$, all theoretical approaches predict a second-order
phase transition, but there are still substantial discrepancies between 
Monte Carlo and three-loop FT calculations 
(see the discussion of Ref.~\cite{LSDASD-00} for $N=6$). 

All these considerations show that a satisfactory theoretical understanding
has not yet been reached. It is not clear whether experiments are observing
first-order transitions in disguise or 
field theory is unable to describe these rather complex systems.
Of course, one may think that the observed disagreement is due to the 
shortness of the available
series, thereby calling for an extension of the perturbative 
expansions to clarify the issue.

FT studies of systems with noncollinear order are based on the
O($N$)$\times$O($M$) symmetric  Hamiltonian~\cite{Kawamura-88,Kawamura-98}
\begin{eqnarray}
{\cal H} = \int d^d x 
&& \left\{ {1\over2}
      \sum_{a} \left[ (\partial_\mu \phi_{a})^2 + r \phi_{a}^2 \right] 
+ {1\over 4!}u_0 \left( \sum_a \phi_a^2\right)^2 \right. \nonumber\\
&& \left. + {1\over 4!}  v_0 
\sum_{a,b} \left[ ( \phi_a \cdot \phi_b)^2 - \phi_a^2\phi_b^2\right]
             \right\},\label{LGWH}
\end{eqnarray}
where $\phi_a$ ($1\leq a\leq M$) are $M$ sets of $N$-component vectors.
We will consider the case $M=2$, that, for $v_0>0$,
describes frustrated systems with
noncollinear ordering such as STA's.
Negative values of $v_0$ correspond to simple ferromagnetic or 
antiferromagnetic 
ordering, and to magnets with sinusoidal spin structures~\cite{Kawamura-88}.

For $N=2$, which is the case relevant for 
frustrated two-component spin models,
an $\epsilon$-expansion analysis indicates
the presence of four fixed points:
the Gaussian one,
an $XY$ fixed point,
an O(4)-symmetric and a mixed fixed point. 
Using nonperturbative arguments~\cite{CB-78}, 
one can show that the $XY$ fixed point
is the only stable one~\cite{alpha} among them.
However, the region relevant for frustrated models, $v_0>0$, is outside 
the domain of attraction of the $XY$ fixed point, which would imply 
a first-order transition.
However, it is still possible that other fixed points
are present in the region $v_0>0$, although they are not predicted
by the $\epsilon$-expansion. 
For $N=3$, one may easily show the existence of 
an O(6) fixed point for $v_0=0$, which is expected 
to be unstable~\cite{Kawamura-98}. 
According to the three-loop analyses of 
Refs.~\cite{AS-94,ASV-95} no other fixed points are 
found for $N=3$, which would imply that the transition is of first order
as well.

In order to investigate the existence of new fixed points, 
we have considered the fixed-dimension perturbative approach,
extending  
the three-loop series of Ref.~\cite{AS-94} to six loops.
As we shall see, the results of our six-loop analysis are
somehow surprising, contradicting most of the earlier FT works.
Indeed, the analysis of the longer series provides a rather robust evidence 
for the existence of a new stable fixed point in the 
$XY$ and Heisenberg cases, with critical exponents that are in 
agreement with the experimental results.

\begin{table}[tbp]
\squeezetable
\caption{
Coefficients $b^{(u)}_{ij}$ 
of the six-loop expansion of 
$\beta_{\bar{u}} =
- \bar{u} + \sum_{i+j \geq 2} b^{(u)}_{ij} \bar{u}^i \bar{v}^j$.
}
\label{betauc}
\renewcommand\arraystretch{0.3}
\begin{tabular}{cl}
\multicolumn{1}{c}{$i,j$}&
\multicolumn{1}{c}{$R_{2N}^{1-i} b^{(u)}_{ij}$}\\
\tableline \hline
2,0 &$ R_{2N}^{-1}$ \\
1,1 &$ -2(N-1)/9 $ \\
0,2 &$ (N-1)/9$\\ \hline
3,0 &$ -8(95+41N)/2187 $ \\
2,1 &$ 400 (N-1) /2187 $ \\
1,2 &$ -118 ( N-1 )/729 $\\
0,3 &$ 10 (N-1)/ 243 $\\ \hline
4,0 &$ 0.27385517 + 0.15072806\,N + 0.0074016064\,{N^2}$\\
3,1 &$ ( N-1 )\,( -0.22580775 - 0.018235606\,N ) $\\
2,2 &$ ( N-1 ) \,( 0.26899935 + 0.020014607\,N ) $\\
1,3 &$ ( N-1 )\,( -0.11448007 - 0.0052835205\,N ) $\\
0,4 &$ ( N-1 )\,( 0.0089444441 - 0.0016624747\,N )$\\\hline
5,0 &$ -0.2792572 - 0.1836675\,N - 0.021838259\,{N^2} $\\&$
+ 0.00018978314\,{N^3}$\\
4,1 &$ ( N-1 ) \,
   ( 0.31461221 + 0.055501872\,N - 0.00082480232\,{N^2} ) $\\
3,2 &$ ( N-1 ) \,
   ( -0.44030468 - 0.073595414\,N + 0.0014028798\,{N^2} ) $\\
2,3 &$ ( N-1 ) \,
   ( 0.2585304 + 0.037648759\,N - 0.001217462\,{N^2} ) $\\
1,4 &$ ( N-1 ) \,
   ( -0.059220508 - 0.0045473385\,N + 0.00073177586\,{N^2} ) $\\
0,5 &$ ( N -1 ) \,
   ( 0.0068557417 - 0.00025180575\,N - 0.00025443043\,{N^2} )$\\\hline
6,0 &$ 0.35174477 + 0.26485003\,N + 0.045288106\,{N^2} $\\&$
+ 0.00043866975\,{N^3} + 0.000013883029\,{N^4}$\\
5,1 &$ ( N-1 ) \,
   ( -0.50696692 - 0.12967024\,N - 0.0014771485\,{N^2} $\\&$
-      0.000084746086\,{N^3} ) $\\
4,2 &$ ( N-1 ) \,
   ( 0.80562212 + 0.20236453\,N + 0.0017578027\,{N^2} $\\&$
+ 0.00018710339\,{N^3} ) $\\
3,3 &$ ( N-1 ) \,
   ( -0.59063997 - 0.14164236\,N - 0.00017117513\,{N^2} 
$\\&$- 0.00019649398\,{N^3} ) $\\
2,4 &$ ( N-1 ) \,
   ( 0.21149268 + 0.045057077\,N - 0.00090030169\,{N^2} 
$\\&$+ 0.000094681347\,{N^3} ) $\\
1,5 &$ ( N-1 ) \,
   ( -0.040302604 - 0.0073126947\,N + 0.00040413932\,{N^2} 
$\\&$- 1.0675626\,\times{{10}^{-6}}\,{N^3} ) $\\
0,6 &$ ( N-1 ) \,
   ( 0.0021121351 + 0.00061386793\,N + 0.000014729631\,{N^2} 
$\\&$- 0.000012344978\,{N^3} ) $\\\hline
7,0 &$ -0.5104989 - 0.4297050\,N - 0.09535750\,{N^2} 
$\\&$ - 0.004001735\,{N^3} + 
   0.00003226842\,{N^4} + 1.410456\,\times{{10}^{-6}}\,{N^5}$\\
6,1 &$ ( N-1 ) \,
   ( 0.8994695 + 0.3004550\,N + 0.01442862\,{N^2} 
$\\&$- 
     0.0001545778\,{N^3} - 0.00001094158\,{N^4} ) $\\
5,2 &$ ( N-1 ) \,
   ( -1.571308 - 0.5310390\,N - 0.02432978\,{N^2} $\\&$+ 
     0.0002518739\,{N^3} + 0.00002911732\,{N^4} ) $\\
4,3 &$ ( N-1 ) \,
   ( 1.354181 + 0.4660641\,N + 0.01959649\,{N^2} $\\&$- 
     0.0001820084\,{N^3} - 0.00003790796\,{N^4} ) $\\
3,4 &$ ( N-1 ) \,
   ( -0.6451180 - 0.2271141\,N - 0.007968847\,{N^2} $\\&$+ 
     0.00009326566\,{N^3} + 0.00002730379\,{N^4} ) $\\
2,5 &$ ( N-1 ) \,
   ( 0.1858305 + 0.0666827\,N + 0.001642878\,{N^2} $\\&$- 
     0.00006134290\,{N^3} - 0.00001228467\,{N^4} ) $\\
1,6 &$ ( N-1 ) \,
   ( -0.02753284 - 0.009832742\,N - 0.00004985576\,{N^2} $\\&$+ 
     0.00001827182\,{N^3} + 4.881350\,\times{{10}^{-6}}\,{N^4} ) $\\
0,7 &$ ( N-1)\,
   ( 0.002029855 + 0.0003424570\,N - 0.00005976447\,{N^2} $\\&$+ 
     3.644114\,\times {{10}^{-6}}\,{N^3} - 1.527023\,\times {{10}^{-6}}\,{N^4} ) $\\
\end{tabular}
\end{table}

\begin{table}[tbp]
\squeezetable
\caption{
Coefficients $b^{(v)}_{ij}$ of the
six-loop expansion of 
$\beta_{\bar{v}} = 
\bar{v} ( - 1 + \sum_{i+j \geq 1} b^{(v)}_{ij} \bar{u}^i \bar{v}^j)$.
}
\label{betavc}
\renewcommand\arraystretch{0.3}
\begin{tabular}{cl}
\multicolumn{1}{c}{$i,j$}&
\multicolumn{1}{c}{$R_{2N}^{-i} b^{(v)}_{ij}$}\\
\tableline \hline
1,0  &$ 4/3 $ \\ 
0,1  &$ (N-6)/9 $ \\\hline
2,0  &$ - 8 (185+23N)/2187 $ \\
1,1  &$ 8 (139 - 4N)/2187 $ \\
0,2  &$ - (86-26N)/729 $ \\\hline
3,0 &$ 0.64380517 + 0.11482552\,N - 0.0068647863\,{N^2}$\\
2,1 &$ -0.65001599 - 0.058188994\,N + 0.014412109\,{N^2}$\\
1,2 &$ 0.25727415 + 0.0050681184\,N - 0.0053156417\,{N^2}$\\
0,3 &$ -0.026488854 + 0.0013330934\,N - 0.00053805835\,{N^2}$\\\hline
4,0 &$ -0.76706177 - 0.17810933\,N + 0.00016284548\,{N^2} $\\&$- 
   0.00070068894\,{N^3}$\\
3,1 &$ 0.95092507 + 0.14711983\,N - 0.013730732\,{N^2} $\\&$+ 
  0.001856844\,{N^3}$\\
2,2 &$ -0.55504891 - 0.065767452\,N + 0.0088371199\,{N^2} $\\&$- 
   0.0016981042\,{N^3}$\\
1,3 &$ 0.14198965 + 0.01319996\,N - 0.0006025682\,{N^2} $\\&$+ 
   0.00067387985\,{N^3}$\\
0,4 &$ -0.014727547 + 0.0017404281\,N + 0.00015290902\,{N^2} $\\&$- 
   0.00011202681\,{N^3}$\\\hline
5,0 &$ 1.096535 + 0.31582586\,N + 0.0094338525\,{N^2} $\\&$- 
   0.00049177077\,{N^3} - 0.000086193996\,{N^4}$\\
4,1 &$ -1.57528 - 0.37431063\,N + 0.0035269473\,{N^2} $\\&$+ 
   0.00032481932\,{N^3} + 0.00027008406\,{N^4}$\\
3,2 &$ 1.1627537 + 0.27564336\,N + 0.0023711857\,{N^2} $\\&$+ 
   0.0008616286\,{N^3} - 0.00032357281\,{N^4}$\\
2,3 &$ -0.44954276 - 0.11599535\,N - 0.0051299898\,{N^2} $\\&$- 
   0.00069713911\,{N^3} + 0.00018853078\,{N^4}$\\
1,4 &$ 0.087684848 + 0.020751579\,N - 0.000084977205\,{N^2} $\\&$- 
   0.000048067825\,{N^3} - 0.000053679999\,{N^4}$\\
0,5 &$ -0.0057616867 - 0.0020152992\,N + 0.0002711957\,{N^2} $\\&$+ 
   0.00010532804\,{N^3} + 5.5607524\,{{10}^{-6}}\,{N^4}$\\\hline
6,0 &$ -1.774553 - 0.6080863\,N - 0.03773523\,{N^2} $\\&$+ 
   0.0005359509\,{N^3} - 0.0001051598\,{N^4} - 0.00001200997\,{N^5}$\\
5,1 &$ 2.864540 + 0.9077083\,N + 0.03756776\,{N^2} $\\&$- 
   0.001936546\,{N^3} + 0.0002352010\,{N^4} + 0.00004342485\,{N^5}$\\
4,2 &$ -2.531676 - 0.8792677\,N - 0.05567215\,{N^2} $\\&$+ 
   0.001133243\,{N^3} - 0.0001483075\,{N^4} - 0.00006347444\,{N^5}$\\
3,3 &$ 1.301582 + 0.5152616\,N + 0.04259343\,{N^2} $\\&$- 
   0.001285064\,{N^3} - 1.472629\,{{10}^{-6}}\,{N^4} + 
   0.00004843683\,{N^5}$\\
2,4 &$ -0.3936964 - 0.1615155\,N - 0.01003169\,{N^2} $\\&$+ 
   0.001578075\,{N^3} + 0.00003181647\,{N^4} - 0.00002045272\,{N^5}$\\
1,5 &$ 0.06196875 + 0.02488608\,N + 0.0004109118\,{N^2} $\\&$- 
   0.0005760149\,{N^3} - 0.00001482553\,{N^4} + 
   4.550213\,{{10}^{-6}}\,{N^5}$\\
0,6 &$ -0.004223551 - 0.0009411514\,N - 0.0000154567\,{N^2} $\\&$+ 
   0.0000215354\,{N^3} + 4.496022\,{{10}^{-6}}\,{N^4} - 
   4.375207\,{{10}^{-7}}\,{N^5}$\\
\end{tabular}
\end{table}

In the fixed-dimension FT approach 
one expands in powers of the 
quartic couplings and renormalizes the theory
by introducing a set of zero-momentum conditions 
for the two-point and four-point correlation 
functions. All perturbative series are finally expressed in terms 
of the zero-momentum four-point renormalized couplings $u$ and $v$ 
normalized so that, at tree level, $u\approx u_0$ and $v\approx v_0$.
The fixed points of the theory are given by 
the common  zeros of the $\beta$-functions $\beta_u(u,v)$ and $\beta_v(u,v)$.
In the case of a continuous transition,
when $m\rightarrow 0$, the couplings $u,v$ are driven toward an
infrared-stable zero $u^*,v^*$ of the  $\beta$-functions.
On the other hand, the absence of stable fixed points is usually considered 
as an indication of a  (weak) first-order transition.

In Tables~\ref{betauc} and \ref{betavc} we present the 
six-loop expansion of the $\beta$-functions for $M=2$,
associated respectively with the rescaled couplings
$\bar{u} = 3 u/( 16 \pi R_{2N})$ and $\bar{v}= 3 v/(16\pi)$, 
where $R_K\equiv 9/(8+K)$.
Since FT perturbative expansions are asymptotic, 
the resummation of the series is essential
to obtain accurate estimates of the physical quantities.
For this purpose
we studied the large-order behavior of the 
expansion in $\bar{u}$ and $\bar{v}$ at 
fixed $z \equiv  \bar{v}/\bar{u}$.
For $z \equiv \bar{v}/\bar{u}$ fixed and $M=2$, 
the singularity of the Borel transform 
closest to the origin, $\bar{u}_b$, is given by 
\begin{eqnarray}
{1\over \bar{u}_b} &=& - a R_{2N}\qquad {\rm for} \;\; 4R_{2N}>z > 0 ,\label{bsing}\\
{1\over \bar{u}_b} &=& - a \left( R_{2N} - {1\over 2} z\right)
\qquad {\rm for} \;\;  z < 0, \;\;\;z > 4R_{2N}, \nonumber
\end{eqnarray}
where $a = 0.14777422...$ and $R_K=9/(8+K)$.
Moreover, we find that for $z> 2 R_{2N}$ the Borel transform 
has a singularity on the positive real axis, which however 
is not the closest one for $z < 4R_{2N}$. Thus, for $z> 2 R_{2N}$
the series is not Borel summable.

\begin{figure}
\vspace*{-4truecm} \hspace*{-3cm}
\epsfxsize = 0.95\textwidth
\leavevmode\epsffile{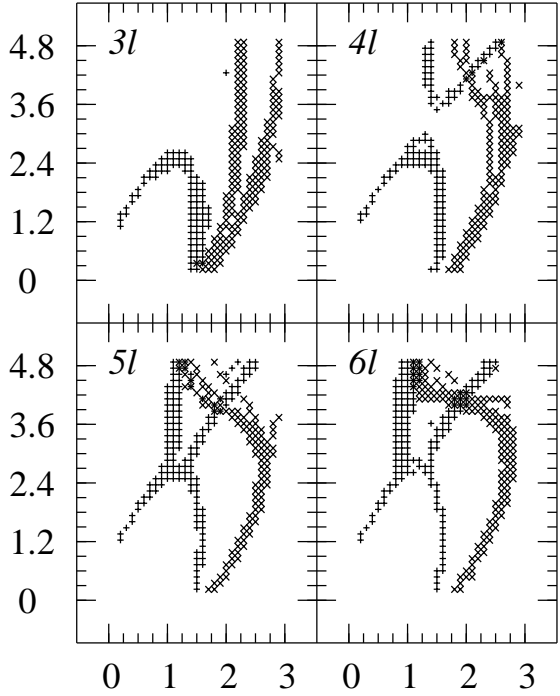}
\vspace*{-9.2truecm}
\caption{
Zeroes of the $\beta$-functions for $N=2$ in the $({\bar u},{\bar v})$ plane.
Pluses ($+$) and crosses ($\times$) correspond to zeros 
of $\beta_{\bar u}({\bar u},{\bar v})$ and 
   $\beta_{\bar v}(\bar{u},\bar{v})$ respectively.
}
\label{fign2}
\end{figure}

\begin{figure}
\vspace*{-4truecm} \hspace*{-3cm}
\epsfxsize = 0.95\textwidth
\leavevmode\epsffile{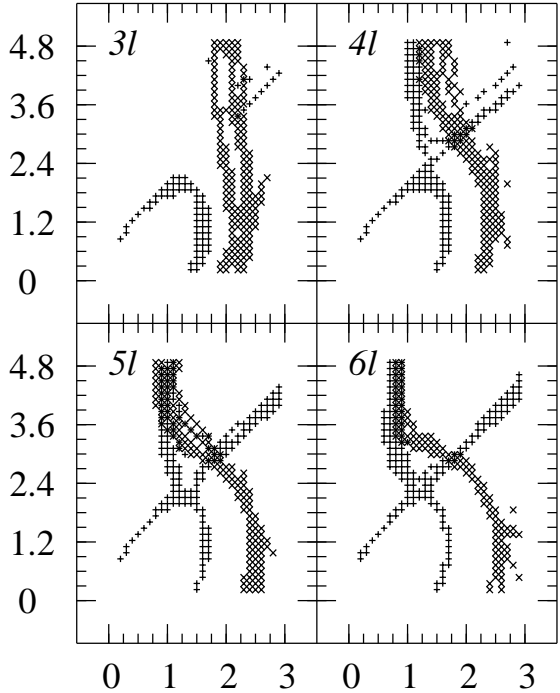}
\vspace*{-9.2truecm}
\caption{
Zeroes of the $\beta$-functions for $N=3$ in the $(\bar{u},\bar{v})$ plane.
Pluses ($+$) and crosses ($\times$) correspond to zeros 
of $\beta_{\bar u}(\bar{u},\bar{v})$ and 
   $\beta_{\bar v}({\bar u},\bar{v})$ respectively.
}
\label{fign3}
\end{figure}

In order to determine the fixed points we use the same method applied
in Ref.~\cite{CPV-00} to the analysis of
the RG functions  of the cubic model.
We resum the perturbative series by means of an 
appropriate  conformal mapping \cite{LZ-77} that takes into
account the large-order behavior of the perturbative series at fixed
$z$ and turns the original series into a convergent sequence of 
approximations. To understand the systematic errors
we vary two different parameters, $b$ and $\alpha$, in the analysis.
We apply this method also for those values of $z$ for which the
series is not Borel summable. Although in this case the sequence of 
approximations is only
asymptotic, it should provide reasonable estimates as long as 
$z < 4 R_{2N}$, since we are taking into account the 
leading large-order behavior. 

In Figs. \ref{fign2} and \ref{fign3} we report our results for the 
zeros of the $\beta$-functions, obtained from the analysis of the $l$-loop
series, $l=3,4,5,6$. 
For each $\beta$-function we consider 18 different approximants 
with $b=3,6,\ldots,18$ and $\alpha=0,2,4$ and we determine the lines
in the $(\bar{u},\bar{v})$ plane on which they vanish. Then, we divide 
the domain $0\le \bar{u}\le 4$ and $0\le \bar{v} \le 5$ into 
$40^2$ rectangles, marking 
those in which at least {\em two} approximants of each $\beta$ function
vanish. No fixed point is observed at 
three loops, consistently with Ref. \cite{AS-94}.
As the number of loops increases, a new fixed point---quite stable 
with respect to the number of loops---clearly appears for
\begin{eqnarray}
\bar{u}^* = 1.9(1), \qquad \bar{v}^* = 4.10(15), && 
  \qquad {\rm for}\;\; N=2, \\
\bar{u}^* = 1.8(1), \qquad \bar{v}^* = 3.00(15), && 
  \qquad {\rm for}\;\; N=3.
\end{eqnarray}
We have been conservative in setting the error bars:
all zeros of the approximants with $3\le b \le 18$ and $0\le \alpha\le 4$
lie within the reported confidence interval. Notice that the fixed points
belong to the region in which the series are not Borel summable,
but still satisfy $\bar{v}^*/\bar{u}^* < 4 R_{2N}$. Therefore, we expect
our resummations to be reliable, and the stability of the results 
with respect to the order of the series confirms it.

We then compute the eigenvalues of the stability matrix. 
They vary significantly with the two parameters
$\alpha$ and $b$ and turn out to be
complex in most of the cases.
Nonetheless, the sign of the real part of the eigenvalues
is always positive showing the stability of the new fixed points.
A reasonable estimate of the exponent $\omega$ is however impossible.
\begin{table}[tbp]
\squeezetable
\caption{
Critical exponents for $N=2$ and $N=3$. Our results are labelled by FT.
Experimental and Monte Carlo results
are reviewed, e.g., in Ref.~\protect\cite{Kawamura-98}. See also
Refs. \protect\cite{MC2}.
}
\label{expexp}
\begin{tabular}{llcccc}
\multicolumn{1}{c}{$N$}&
\multicolumn{1}{c}{}&
\multicolumn{1}{c}{$\gamma$}&
\multicolumn{1}{c}{$\nu$}&
\multicolumn{1}{c}{$\beta$}&
\multicolumn{1}{c}{$\alpha$}\\
\tableline \hline
$2$ & CsMnBr$_3$ & 1.10(5) \cite{ref25} & 0.57(3) \cite{ref25} 
    & 0.25(1) \cite{ref25} & 0.39(9) \cite{ref26}\\
      &            & 1.01(8) \cite{ref24} & 0.54(3) \cite{ref24} & 0.22(2) \cite{ref24} & 0.40(5)\cite{ref27} \\
      &            & & & 0.24(2) \cite{ref39}  &  \\
      & CsNiCl$_3$ & & & 0.243(5) \cite{ref93} & 0.37(8)\cite{ref91}   \\
      &            & & & &  0.342(5)\cite{ref92}   \\
      & CsMnI$_3$ & & & & 0.34(6)\cite{ref91}  \\
      & MC \cite{Kawamura-92} & 1.13(5) & 0.54(2) & 0.253(10) & 0.34(6) \\
      & FT  & 1.10(4) & 0.57(3) & 0.31(2) & 0.29(9) \\\hline
$3$ & VCl$_2$   & 1.05(3) \cite{ref29} & 0.62(5) \cite{ref29} & 0.20(2) \cite{ref29} &   \\
      & VBr$_2$   & & & & 0.30(5) \cite{ref30}   \\
      & RbNiCl$_3$& & & 0.28(1) \cite{ref95} &  \\
      & CsNiCl$_3$& & & 0.28(3) \cite{ref93} & 0.25(8) \cite{ref91}  \\
      & & & & & 0.23(4) \cite{ref92}  \\
      & & & & & 0.28(6) \cite{ref138}  \\
      & MC \cite{Kawamura-92} & 1.17(7) & 0.59(2) & 0.30(2) & 0.24(8) \\
      & FT & 1.06(5) & 0.55(3) & 0.30(2) & 0.35(9) \\
\end{tabular}
\end{table}
We also compute the critical exponents, by estimating the corresponding
six-loops series at the fixed point, following 
Ref. \cite{CPV-00}. The results are in substantial
agreement with the experimental and Monte Carlo (MC)
estimates, see Table \ref{expexp}.

Finally, we compare the six-loop results with 
the critical exponents computed to $O(1/N^2)$ in the framework of
the large-$N$ expansion. For example,
\begin{eqnarray}
&&\nu = 1 - {16\over \pi^2} {1\over N} - \left( {56\over \pi^2}-{640\over 3 \pi^4} \right) {1\over N^2}
+O\left( {1\over N^3}\right).\label{nuln}
\end{eqnarray}
We find $\nu=0.858(4)$ for $N=16$ and $\nu=0.936(2)$
for $N=32$, which compare reasonably with the estimates that one obtains from
Eq.~(\ref{nuln}), i.e. $\nu=0.885$ for $N=16$ and $\nu=0.946$ for $N=32$.

In conclusion, the extension to six loops of the FT expansions
solves the apparent contradictions between field theory and 
experiments. We find that new stable chiral fixed points exist
for two- and three-component systems. The estimated exponents are
in substantial agreement with experiments, whose conclusions 
on the nature of the phase transitions are thus confirmed. 
However, we note that 
first-order transitions are still possible for systems that 
are outside the attraction domain of the chiral fixed point.
In this case, the RG flow runs away to a 
first-order transition.


\begin{references}

\bibitem{Kawamura-98}
H.~Kawamura, J. Phys.: Condens. Matter {\bf 10}, 4707 (1998).

\bibitem{AS-94}
S.~A.~Antonenko and A.~I.~Sokolov,
Phys. Rev. B {\bf 49}, 15901 (1994).

\bibitem{ASV-95}
S.~A.~Antonenko, A.~I.~Sokolov and V.~B.~Varnashev,
Phys. Lett. A {\bf 208}, 161 (1995).

\bibitem{ERG}
G. Zumbach,  Nucl. Phys. B {\bf 413}, 771 (1994).
M.~Tissier, B.~Delamotte, and D.~Mouhanna,
cond-mat/0001350.

\bibitem{MC}
D.~Loison and K.~D.~Schotte,
Eur. Phys. J. B {\bf 5}, 735 (1998);
cond-mat/0001135. 

\bibitem{LSDASD-00}
D.~Loison et al.,
cond-mat/0001105.

\bibitem{Kawamura-88}
H.~Kawamura, Phys. Rev. B {\bf 38}, 4916 (1988);
erratum  B {\bf 42}, 2610 (1990).

\bibitem{CB-78}
R. A. Cowley and A. D. Bruce,
J. Phys. C: Solid State Phys. {\bf 11}, 3577 (1978).


\bibitem{alpha}
This fact is confirmed by the analysis of the six-loop expansion, 
which also show that the O(4) and mixed fixed points are unstable.

\bibitem{CPV-00}
J.~M. Carmona, A. Pelissetto, and E. Vicari,
Phys. Rev. B {\bf 61}, 15136 (2000).

\bibitem{LZ-77} J.~C.~Le Guillou and J.~Zinn-Justin,
Phys.\ Rev.\ B {\bf 21}, 3976 (1980). 

\bibitem{MC2}
T. Bhattacharya et al., J. Phys. I (Paris) {\bf 4}, 181 (1994).
A.~Mailhot at al., Phys. Rev. B {\bf 50}, 6854 (1994).
M.~L.~Plumer and A.~Mailhot, Phys. Rev. B {\bf 50}, 16113 (1994).
D.~Loison and H.~T.~Diep,  Phys. Rev. B {\bf 50}, 16453 (1994).
E.~H.~Boubcheur et al., Phys. Rev. B {\bf 54}, 4165 (1996).

\bibitem{ref25}
H.~Kadowaki et al.,
J. Phys. Soc. Japan {\bf 57}, 2640 (1988).
Y. Ajiro et al.,
J. Phys. Soc. Japan {\bf 57}, 2648 (1988).

\bibitem{ref26}
J.~Wang et al.,
Phys. Rev. Lett. {\bf 66}, 3195 (1991).

\bibitem{ref24}
T. E. Mason et al.,
J. Phys. C: Solid State Phys. {\bf 20}, L945 (1987);
Phys. Rev. B {\bf 39}, 586 (1989).

\bibitem{ref27}
R.~Deutschmann et al., 
Europhys. Lett. {\bf 17}, 637 (1992).

\bibitem{ref39}
B. D. Gaulin et al.,
Phys. Rev. Lett. {\bf 62}, 1380 (1989).

\bibitem{ref93}
M. Enderle et al.,
Physica B {\bf 234-236}, 554 (1997).

\bibitem{ref91}
D. Beckmann et al.,
Phys. Rev. Lett. {\bf 71}, 2829 (1993).

\bibitem{ref92}
M. Enderle et al.,
J. Phys.: Condens. Matter {\bf 6}, L385 (1994).

\bibitem{Kawamura-92}
H. Kawamura, J. Phys. Soc. Japan {\bf 61}, 1299 (1992).

\bibitem{ref29}
H. Kadowaki et al.,
J. Phys. Soc. Japan {\bf 56}, 4027 (1988).

\bibitem{ref30}
J. Wosnitza et al.,
J. Phys.: Condens. Matter {\bf 6}, 8045 (1994).

\bibitem{ref95}
Y. Oohara et al.,
J. Phys. Soc. Japan {\bf 60}, 393 (1991).

\bibitem{ref138}
H. Weber et al.,
Int. J. Mod. Phys. B {\bf 9}, 1387 (1995).





\end{references}
\end{document}